\documentclass[global,twocolumn,referee]{svjour}
\usepackage{graphics}
\usepackage{amsmath,amssymb}
\journalname{ar$\chi$iv}
\begin{document}
\title{The direct evaluation of attosecond chirp from a streaking measurement}
\author{Justin Gagnon\inst{1,2} \and Vladislav S. Yakovlev\inst{1,2}
}                     
%
%
\institute{Max-Planck-Institut f\"{u}r Quantenoptik, Hans-Kopfermann-Str. 1, D-85748 Garching, Germany \and Ludwig-Maximilians-Universit\"{a}t M\"{u}nchen, \\ Am Coulombwall 1,D-85748 Garching, Germany\\\email{justin.gagnon@mpq.mpg.de}}

\date{Received: date / Revised version: date}
%
\maketitle
\begin{abstract}
We derive an analytical expression that relates the breadth of a streaked photoelectron spectrum to the group-delay dispersion of an isolated attosecond pulse. Based on this analytical expression, we introduce a simple, efficient and robust procedure to instantly extract the attosecond pulse's chirp from the streaking measurement. We show that our method is robust against experimental artifacts.
\end{abstract}
\section{Introduction}
\label{intro}
The characterization of isolated attosecond pulses has played an important role in the development of attosecond science \cite{Hentschel:2001,Baltuska:2003,Krausz:2009}. The generation and application of ever shorter attosecond extreme-ultraviolet (XUV)
pulses \cite{Nisoli:2006,Goulielmakis:2008} relies on knowledge of their time-domain properties, which can be obtained by means of attosecond streaking measurements \cite{Itatani:2002,Kienberger:2004}.

So far, the main functions of attosecond streaking are (i) to characterize the field of an attosecond pulse and (ii) to temporally resolve a physical process on the attosecond scale. Here, we are concerned with the former application of attosecond streaking, that of characterizing an attosecond pulse. Much effort has been exerted on the development of methods for extracting physical information from the streaking measurement \cite{Mairesse:2005,Ge:2008,Chini:2010}, with the current state-of-the-art being the FROG retrieval algorithm \cite{Trebino:1997}. The FROG algorithm has already been used to characterize the shortest attosecond pulses \cite{Goulielmakis:2008}, and to uncover a measured delay of $20\,\textrm{as}$ between photoemissions from the $2s$ and $2p$ sub-shells of neon \cite{Schultze:2010}. It is relatively robust \cite{Gagnon:2009}, and provides a wealth of information about the temporal characteristics of the attosecond and laser fields \cite{Gagnon:2008}.

However, the application of FROG to attosecond streaking requires quite stringent experimental requirements, such as a sufficient amount of recorded spectra, with a delay step between them on the order of the attosecond pulse's duration. These experimental parameters become unwieldy as the duration of attosecond pulses approaches the atomic unit of time. Moreover, the FROG algorithm is a somewhat complicated numerical optimization procedure, whose output (the attosecond field and the laser field) is not transparently related to the input (the set of streaked spectra). Thus, errors in the reconstructed pulses are difficult to interpret due to the FROG algorithm's black-box nature. Although FROG provides a complete characterization of the attosecond XUV field, the \emph{duration} of the attosecond pulse is the primordial quantity that will be interrogated as attosecond streaking continues to expand beyond its original scope into various research fields.

In this article, we introduce a simple and robust method for quantifying the chirp of an attosecond pulse based on an analytical formula we derive from laser-dressed photoelectron trajectories. Using this formula, we develop a method that \emph{directly} evaluates the attosecond pulse's group-delay dispersion from a sequence of streaked spectra, which in turn sets the pulse's duration provided its spectrum is known. Our method avoids the stringent experimental conditions required for the attosecond FROG technique, and provides accurate results with very few electron spectra in a matter of seconds. We begin this article with the derivation of the analytical expression for the change in photoelectron bandwidth due to the streaking effect, and then introduce our method with a numerical example. All quantities are expressed in atomic units unless otherwise stated.

\section{Classical electron trajectory analysis of the streaking effect}

Let us first consider an attosecond XUV pulse with electric field $F_\mathrm{X}(t)$ given by
\begin{subequations}
\label{attosecond_field}
\begin{align}
  \label{attosecond_field_field}
  F_\mathrm{X}(t)&=|F_\mathrm{X}(t)|\mathrm{e}^{\mathrm{i}\left(\Omega_\mathrm{X} t+\varphi_\mathrm{X}(t)\right)},\\
  \label{attosecond_chirp}
  \varphi_\mathrm{X}(t)&=\frac{1}{2}\beta_1 t^2+\frac{1}{6}\beta_2 t^3+\ldots,
\end{align}
\end{subequations}
where the spectrum of the attosecond pulse is centered at $\Omega_\mathrm{X}$ with small variations in frequency due to the higher-order temporal phase $\varphi_\mathrm{X}(t)$. The attosecond pulse launches electron trajectories that are parameterized with an initial time $t$ as well as an electron energy $\varepsilon=p^2/2$. Due to the attosecond pulse's finite \emph{bandwidth}, we consider the energy $\varepsilon$ as an independent variable, while the independent variable $t$ is a result of the finite \emph{duration} of the attosecond pulse. Thus, the set of trajectories is described by a time-energy distribution with respect to $\{t,\varepsilon\}$.

The final energy $\varepsilon_\mathrm{S}$ of an electron, launched at some moment $t$ in a continuum permeated by a near-infrared (NIR) laser field, is then
\begin{subequations}
  \begin{align}
    \label{final_energy}
    \varepsilon_\mathrm{S}&=\frac{1}{2}\left(\sqrt{2\big(\varepsilon+\omega_\mathrm{X}(t)\big)}-A_\mathrm{L}(t)\right)^2\\
    \label{final_energy_approx}
    &\approx\varepsilon-p A_\mathrm{L}(t)+\frac{1}{2}A_\mathrm{L}^2(t)+\left(1-\frac{A_\mathrm{L}(t)}{p}\right)\omega_\mathrm{X}(t),
  \end{align}
\end{subequations}
where we define the instantaneous frequency $\omega_\mathrm{X}(t)=\dot{\varphi}_\mathrm{X}(t)$ due to the chirp of the attosecond pulse, and $A_\mathrm{L}(t)$ is the vector potential of the laser field. Since the change in frequency over the temporal profile of the attosecond pulse is much smaller than the central frequency $\Omega_\mathrm{X}$, the last term in (\ref{final_energy_approx}) is comparably small and can be dropped, leading to the simple relation $\varepsilon_\mathrm{S}\approx\varepsilon-p A_\mathrm{L}(t)+A_\mathrm{L}^2(t)/2$ for the shift of the photoelectron spectrum.

It is known \cite{Kienberger:2004} that the spectral shift alone is not sufficient to obtain information about the attosecond pulse's chirp because the final energy $\varepsilon_\mathrm{S}$ is hardly sensitive to the temporal phase $\varphi_\mathrm{X}(t)$ of the attosecond pulse. The main manifestation of the attosecond chirp in the streaking measurement is the change in breadth of the streaked photoelectron spectrum. To describe this effect, we interpret (\ref{final_energy}) as a mapping of the initial time and energy of an electron trajectory to a final energy (e.g. measured at the detector). To describe the effect of chirp, it is useful to consider small changes $\mathrm{d}\varepsilon_\mathrm{S}$ in the final energy with respect to small changes in the initial energy $\mathrm{d}\varepsilon$ and time $\mathrm{d}t$ of the trajectory. The total differential of (\ref{final_energy}) is then
\begin{align}
  \label{total_differential}
  \mathrm{d}\varepsilon_\mathrm{S}&\approx\left(1-\frac{A_\mathrm{L}(t)}{p}\right)\Big(\left(\beta_\mathrm{X}(t)+pF_\mathrm{L}(t)\right)\mathrm{d}t+\mathrm{d}\varepsilon\Big),
\end{align}
where we again neglect the small terms containing $\omega_\mathrm{X}(t)$. The temporal phase of the attosecond pulse appears in (\ref{total_differential}) as $\beta_\mathrm{X}(t)=\ddot{\varphi}_\mathrm{X}(t)$, which defines the \emph{chirp} of the attosecond pulse. We have also introduced the electric field of the laser pulse $F_\mathrm{L}(t)=-\dot{A}_\mathrm{L}(t)$. Thus, the chirp of the attosecond pulse and the electric field of the laser pulse both influence the spread in final energies resulting from the streaking effect.

To proceed further, we interpret the effects of the NIR field on the time-energy distribution of electron trajectories, as described by (\ref{total_differential}), in a straightforward manner. Initial inspection of (\ref{total_differential}) shows that the NIR field imparts an additional energy sweep of $pF_\mathrm{L}(t)$ to the photoelectron, resulting in a total chirp $\beta_\mathrm{S}(t)=\beta_\mathrm{X}(t)+pF_\mathrm{L}(t)$. Furthermore, the NIR field re-scales the energy spread by a factor $(1-A_\mathrm{L}(t)/p)$. As a result, both the NIR electric field $F_\mathrm{L}(t)$ and the NIR vector potential $A_\mathrm{L}(t)$ have a role in modifying the breadth of the photoelectron spectrum.

In order to account for the effects of the streaking field, we recall that the attosecond electron wave packet can be viewed as a replica \cite{Itatani:2002} of the attosecond pulse $F_\mathrm{X}(t)$. We model this photoelectron replica as
\begin{align}
  \label{unstreaked_electron_wave_packet}
  \chi(t)&=\mathrm{e}^{-\frac{1}{2}(t/\tau_\mathrm{X})^2}\mathrm{e}^{\mathrm{i}\left(\varepsilon_\mathrm{C} t+\frac{1}{2}\beta_\mathrm{X}t^2\right)},
\end{align}
where $\varepsilon_\mathrm{C}$ is the central photoelectron energy. Naturally, since the electron trajectories are launched by the attosecond pulse, the duration $\tau_\mathrm{X}$ of the electron wave packet \cite{Yakovlev:2010} should be nearly the same as that of the attosecond pulse; and as $\chi(t)$ is a replica of $F_\mathrm{X}(t)$, its chirp $\beta_\mathrm{X}$ is the same as that of the attosecond pulse. For simplicity, we assume $\beta_\mathrm{X}$ to be constant and we also assume that the attosecond pulse is shorter than any relevant time scale of the NIR field, so that $F_\mathrm{L}(t)$ and $A_\mathrm{L}(t)$ are evaluated at the central time $t_0$ of the attosecond pulse.

Now, in order to include the effects of the streaking field, we first consider the shift of the photoelectron spectrum due to $A_\mathrm{L}(t_0)$ and the change in bandwidth due to the chirp induced by $F_\mathrm{L}(t_0)$. To this end, we modify the wave packet's central energy $\varepsilon_\mathrm{C}=p^2_\mathrm{C}/2$ and chirp $\beta_\mathrm{X}$ as follows:
\begin{subequations}
  \label{modelling_a_streaked_wave_packet}
\begin{align}
  \label{modification_of_central_energy}
  \varepsilon_\mathrm{C}&\longrightarrow\varepsilon_\mathrm{S}=\varepsilon_\mathrm{C}-p_\mathrm{C}A_\mathrm{L}(t_0)+\frac{1}{2}A^2_\mathrm{L}(t_0)\\
  \label{modification_of_chirp}
  \beta_\mathrm{X}&\longrightarrow\beta_\mathrm{S}=\beta_\mathrm{X}(t)+p_\mathrm{C}F_\mathrm{L}(t).
\end{align}
\end{subequations}
With these substitutions, the \emph{streaked} photoelectron wave packet is modeled as
\begin{align}
  \label{streaked_electron_wave_packet}
  \chi_\mathrm{S}(t)&=\mathrm{e}^{-\frac{1}{2}(t/\tau_\mathrm{X})^2}\mathrm{e}^{\mathrm{i}\left(\varepsilon_\mathrm{S}t+\frac{1}{2}\beta_\mathrm{S}t^2\right)}.
\end{align}
To obtain an expression for the bandwidth of the streaked photoelectron spectrum, we note that the streaked photoelectron spectrum is just a Fourier-transform of $\chi_\mathrm{S}(t)$ \cite{Kitzler:2002}. Since the streaked wave packet is a Gaussian, the Fourier transform of $\chi_\mathrm{S}(t)$ can be carried out analytically, yielding the following expression for the bandwidth of the streaked spectrum:
\begin{subequations}
  \label{streaked_bandwidth}
\begin{align}
  \label{streaked_bandwidth_delta}
  \delta_\mathrm{S}(t_0)&=\frac{\delta_\mathrm{X}}{\eta_\mathrm{X}}\left(1-\frac{A_\mathrm{L}(t_0)}{p_\mathrm{C}}\right)\sqrt{\left(\eta_\mathrm{X}^{(0)}\right)^2+\Big(\delta^2_\mathrm{X}\gamma_\mathrm{S}(t_0)\Big)^2},\\
  \label{streaked_gdd}
  \gamma_\mathrm{S}(t_0)&=\gamma_\mathrm{X}+\left(\frac{\eta_\mathrm{X}}{\delta^2_\mathrm{X}}\right)^2 p_\mathrm{C}F_\mathrm{L}(t_0),
\end{align}
\end{subequations}
where $\delta_\mathrm{X}$ and $\gamma_\mathrm{X}$ represent the bandwidth and group-delay dispersion (GDD)---defined as the second derivative of the spectral phase---of the attosecond pulse. The quantity $\eta_\mathrm{X}=\sqrt{\left(\eta_\mathrm{X}^{(0)}\right)^2+\left(\delta^2_\mathrm{X}\gamma_\mathrm{X}\right)^2}$ is the attosecond pulse's time-bandwidth product, with a Fourier-limited time-bandwidth product $\eta_\mathrm{X}^{(0)}$ ($\eta_\mathrm{X}^{(0)}=1/2$ for a Gaussian spectrum). The quantities $\tau_\mathrm{X}$, $\delta_\mathrm{X}$ and $\delta_\mathrm{S}(t_0)$ are all taken as standard deviations of their respective distributions. According to (\ref{streaked_bandwidth}), $\gamma_\mathrm{X}$ determines the width $\delta_\mathrm{S}(t_0)$ of the streaked spectrum as a function of $t_0$. Provided that the characteristics of the field-free spectrum ($\Omega_\mathrm{X}$, $\delta_\mathrm{X}$ and $\eta_\mathrm{X}^{(0)}$) as well as those of the laser field ($A_\mathrm{L}(t)$ and $F_\mathrm{L}(t)$) are known, $\gamma_\mathrm{X}$ remains the only free parameter.

In writing (\ref{streaked_bandwidth}), we also explicitly included the energy re-scaling pre-factor $(1-A_\mathrm{L}(t_0)/p_\mathrm{C})$. Similar but less general expressions for the streaked photoelectron bandwidth were previously derived in \cite{Itatani:2002,Gagnon:2009} from the semi-classical expression for streaking \cite{Kitzler:2002}. These expressions consider photoionization at the zero-crossing of the vector potential, $A_\mathrm{L}(t_0)=0$, where there is no spectral shift but only a change in spectral bandwidth due to the NIR field. These expressions therefore do not contain the bandwidth re-scaling factor $(1-A_\mathrm{L}(t_0)/p_\mathrm{C})$, which is needed to accurately represent the bandwidth of the streaked spectra at arbitrary delay times $t_0$, when the NIR field simultaneously shifts the photoelectron spectrum and changes its bandwidth.

Although (\ref{streaked_bandwidth}) was deduced assuming a Gaussian wave packet, it actually applies to more general pulse shapes owing to the fact that the relation $\eta_\mathrm{X}^2=\left(\eta_\mathrm{X}^{(0)}\right)^2+\delta^4_\mathrm{X}\gamma_\mathrm{X}^2$ holds for arbitrary spectra with a constant GDD (see Appendix A). The following section presents numerical examples in further support of this claim.

\section{A method to directly extract the attosecond chirp from a set of streaked photoelectron spectra}

Equation (\ref{streaked_bandwidth}) serves as the basis for our method to extract the attosecond chirp from a streaking measurement. Our procedure is very straightforward: we evaluate the first moments ($\varepsilon_\mathrm{S}$) of the streaked spectra to obtain the laser field's vector potential $A_\mathrm{L}(t)$, which in turn gives us the laser's electric field $F_\mathrm{L}(t)$. We also compute a curve $\delta_\mathrm{S}^{(\mathrm{M})}(t_0)$ of standard deviations of the measured streaked spectra as a function of the XUV-NIR delay $t_0$. Lastly, we find the attosecond chirp $\gamma_\mathrm{X}$---the only free parameter in (\ref{streaked_bandwidth})---which minimizes the discrepancy between the widths $\delta_\mathrm{S}^{(\mathrm{M})}(t_0)$ obtained from the set of streaked spectra and those given by the model (\ref{streaked_bandwidth}). To compare these two, we define a figure of merit
\begin{align}
  \label{figure_of_merit}
  M&=\frac{\sum_j\left(\delta_\mathrm{S}(t_j)-\delta_\mathrm{S}^{(\mathrm{M})}(t_j)\right)^2}{\sum_j\Big(\delta_\mathrm{S}(t_j)\Big)^2+\sum_j\Big(\delta_\mathrm{S}^{(\mathrm{M})}(t_j)\Big)^2},
\end{align}
where the sums range over the XUV-NIR delays $t_j$. The goal of our procedure is to find $\gamma_\mathrm{X}$ that best reproduces the measured curve $\delta_\mathrm{X}^{(\mathrm{M})}(t_0)$ according to model (\ref{streaked_bandwidth}).

\begin{figure}[t]
\resizebox{0.45\textwidth}{!}{\includegraphics{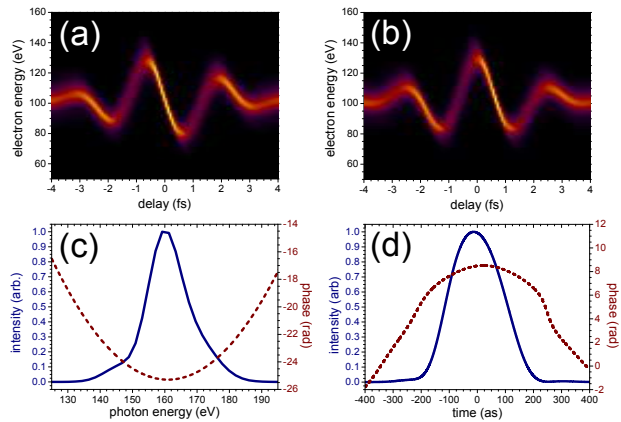}}
\caption{Panels (a) and (b) show sets of 101 streaked photoelectron spectra evaluated by solving the TDSE, using streaking fields with $\phi_0=0$ and $\phi_0=\pi/2$, respectively. Panel (c) shows the attosecond pulse's spectrum (solid line) and phase (dotted line), while panel (d) displays its temporal intensity profile (solid line) and temporal phase (dotted line).}
\label{streaking_example}
\end{figure}

As an example, we consider the case of a non-Gaussian $\sim226\,\textrm{as}$ XUV pulse. This pulse has a constant GDD of $\sim 5885\,\textrm{as}^2$. However, since its spectrum (Figure \ref{streaking_example}-b) is irregular ($\eta_\mathrm{X}^{(0)}\approx 0.5515$), i.e. it is asymmetric and contains some fine structure, its chirp $\beta_\mathrm{X}$ is time-dependent. The streaking field is a NIR pulse given by
\begin{align}
\label{streaking_field}
A_\mathrm{L}(t)&=A_0\cos^4(t/\tau_\mathrm{L})\sin(\omega_\mathrm{L}t+\phi_0)
\end{align}
with $\tau_\mathrm{L}\approx5.743\,\textrm{fs}$, yielding a $3\,\textrm{fs}$ full width at half maximum (FWHM) duration, $\omega_\mathrm{L}\approx2.355\,\textrm{rad/fs}$ corresponding to a central wavelength of $800\,\textrm{nm}$ and with $A_0\approx-0.41915\,\textrm{a.u.}$, giving a peak intensity of $20\,\textrm{TW}/\textrm{cm}^2$. For this example, we consider carrier-envelope phase values of $\phi_0=0$ (Figure \ref{streaking_example}-a) and $\phi_0=\pi/2$ (Figure \ref{streaking_example}-b).

The simulated streaking measurements, shown in Figure \ref{streaking_example}-a and Figure \ref{streaking_example}-b, are composed of a sequence of streaked spectra computed for different delays between the XUV and NIR fields by propagating the time-dependent Schr\"{o}dinger equation (TDSE) using a split-step FFT scheme. The Hamiltonian is that of a single electron in one dimension, assuming a soft-core potential with an ionization energy $W\approx59\,\textrm{eV}$.

\begin{figure}[t]
\resizebox{0.45\textwidth}{!}{\includegraphics{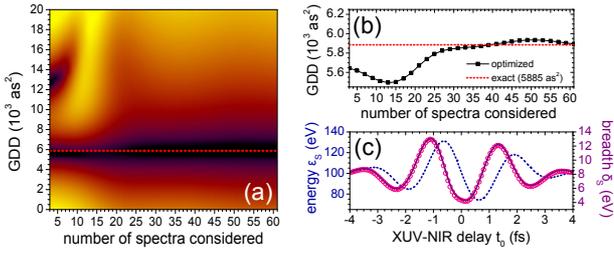}}
\caption{The analytical chirp evaluation (ACE) is applied to the streaking example shown in Figure \ref{streaking_example}-a. Panel (a) is a false-color logarithmic plot of the figure of merit $M$, defined by (\ref{figure_of_merit}), versus the number of spectra ($N$) considered for the ACE procedure. Panel (b) plots the retrieved GDD (squares) at the global minimum of $M$ as a function of $N$. In panels (a) and (b), the dotted red line represents the exact GDD. Panel (c) shows the energy $\varepsilon_\mathrm{S}$ (dotted line) and breadth $\delta_\mathrm{S}^{(\mathrm{M})}(t_0)$ (solid line) evaluated from the streaked spectra. The hollow circles represent the breadths $\delta_\mathrm{S}(t_0)$ computed from (\ref{streaked_bandwidth}) with the exact $\gamma_\mathrm{X}=5885\,\textrm{as}^2$.}
\label{chirp_extraction}
\end{figure}

\begin{figure}[t]
\resizebox{0.45\textwidth}{!}{\includegraphics{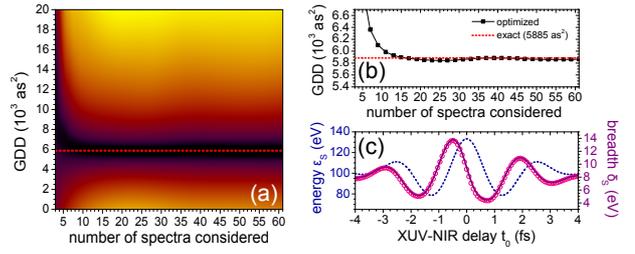}}
\caption{The analytical chirp evaluation (ACE) is applied to the streaking example shown in Figure \ref{streaking_example}-b. The data shown here are presented in the same manner as in Figure \ref{chirp_extraction}.}
\label{chirp_extraction_sine_pulse}
\end{figure}

The results of our analytical chirp evaluation (ACE) procedure, applied to the spectrograms shown in Figures \ref{streaking_example}-a and \ref{streaking_example}-b, are shown in Figures \ref{chirp_extraction} and \ref{chirp_extraction_sine_pulse}. In both cases, we have applied ACE to different subsets of streaked spectra, by considering a varying number $N$ of spectra about the central delay value $t_0=0$.

For the case $\phi_0=0$, Figure \ref{chirp_extraction}-a shows a false-color plot of the figure of merit $M$ as defined in (\ref{figure_of_merit}). Darker areas correspond to a smaller value of $M$. When too few spectra are considered, Figure \ref{chirp_extraction}-a shows a local minimum near $\gamma_\mathrm{X}=12\,500\,\textrm{as}^2$ which disappears as more spectra ($N\gtrsim 13$) are considered. Nonetheless, Figure \ref{chirp_extraction}-b shows that we recover the exact GDD (the dashed line) from the global minimum to within $\sim4\%$ with as few as three spectra. As $N$ increases, the global minimum eventually stabilizes around the red dashed line representing the exact GDD, and ACE converges nearly to the exact value $\gamma_\mathrm{X}=5885\,\textrm{as}^2$. Figure \ref{chirp_extraction}-c shows that the model (\ref{streaked_bandwidth}) reproduces the correct curve $\delta_\mathrm{S}(t_0)$ for the exact GDD. In contrast, we found that the attosecond FROG retrieval \cite{Gagnon:2008} fails to converge when fewer than 25 spectra are included, for which it recovers a GDD $\gamma_\mathrm{X}=5740\,\textrm{as}^2$.

For $\phi_0=\pi/2$, Figure \ref{chirp_extraction_sine_pulse}-ashows that the figure of merit has only one minimum as a function of GDD. This minimum quickly converges to the correct GDD as more spectra are considered in the evaluation, as displayed in Figure \ref{chirp_extraction_sine_pulse}-b, and is already accurate to within $0.7\%$ for $N=13$ spectra. Figure \ref{chirp_extraction_sine_pulse}-c shows that the model (\ref{streaked_bandwidth}) once again reproduces the correct curve (hollow circles) of streaked breadths for the exact GDD $\gamma=5885\,\textrm{as}^2$.

The main advantage of the ACE procedure is that it requires very few spectra. As long as $A_\mathrm{L}(t_0)$ is properly sampled by the delay step between the spectra, there is enough information for ACE to recover the GDD of the attosecond pulse. In contrast, FROG requires the delay step to be on the order of the attosecond pulse's duration. To illustrate this point, we apply ACE to a subset of the spectra shown in Figure \ref{streaking_example}-a and \ref{streaking_example}-b. Specifically, we consider $17$ spectra over the interval  $[-2\,\textrm{fs},1.84\,\textrm{fs}]$ (containing $1.5$ cycles of the streaking field), with a delay step of $240\,\textrm{as}$ between them, i.e. a third of the original spectra in $[-2\,\textrm{fs},2\,\textrm{fs}]$. Even with so few spectra, ACE still recovered accurate GDD's of $6150\,\textrm{as}^2$ and $6110\,\textrm{as}^2$ for $\phi_0=0$ and $\phi_0=\pi/2$, respectively. On the other hand, FROG failed to converge to anything meaningful in both cases, most likely because the delay step was too large.

To further demonstrate ACE's robustness against a non-Gaussian spectrum, we consider a clipped version of the XUV spectrum shown in Figure \ref{streaking_example}-c, for which we remove energy components above $175\,\textrm{eV}$. Experimentally, such a sharp edge in the XUV spectrum might result from the beam's transmission through a metallic filter. Using the clipped XUV spectrum, we compute sets of $101$ streaked photoelectron spectra, with the same parameters as those displayed in Figure \ref{streaking_example}-a and \ref{streaking_example}-b. In spite of this heavy clipping, ACE recovers GDD's of $5940\,\textrm{as}^2$ and $5960\,\textrm{as}^2$ for $\phi_0=0$ and $\phi_0=\pi/2$, respectively. In comparison, FROG recovers a GDD of $5620\,\textrm{as}^2$ for both $\phi_0=0$ and $\phi_0=\pi/2$.

As previously mentioned, these examples assume a constant GDD over an irregular spectral distribution, resulting in a chirp $\beta_\mathrm{X}$ that depends on time. Since expression (\ref{streaked_bandwidth})---which is at the core of the ACE procedure---assumes a constant chirp in time, then the chirp parameter $\beta_\mathrm{X}$ is interpreted as the \emph{average} chirp over the attosecond pulse's temporal profile. Conversely, if a non-uniform GDD was considered, then ACE would have recovered the \emph{average} GDD over the spectral profile.

\begin{figure}[t]
\resizebox{0.45\textwidth}{!}{\includegraphics{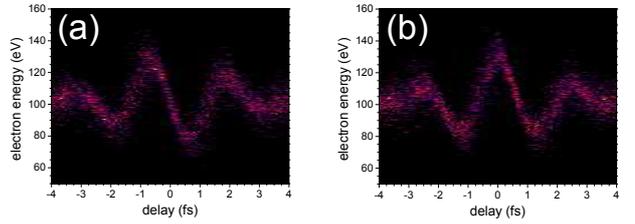}}
\caption{Panels (a) and (b) shows sets of streaked spectra, computed by adding Poisson noise to those of Figures \ref{streaking_example}-a and \ref{streaking_example}-b, respectively. ACE recovers GDD's of $5590\,\textrm{as}^2$ and $6040\,\textrm{as}^2$ from the spectra in panels (a) and (b), respectively.}
\label{noisy_spectrograms}
\end{figure}

As an additional verification of ACE's robustness, we investigate the effect of noise in the streaked spectra. To this end, we add noise to the sets of $101$ spectra shown in Figure \ref{streaking_example}-a and \ref{streaking_example}-b. We assume that the number of counts $n$ in a spectral bin follows a Poisson distribution $P(n;\mu)=\mu^n\mathrm{e}^{-\mu}/n!$, with an expectation value $\mu$ proportional to the spectral intensity (we set $\mu=1$ for the peak of the spectrogram, corresponding to a very low count rate). From these considerations, we compute the noisy spectra which are shown in Figure \ref{noisy_spectrograms}.

Even under such nefarious conditions, ACE recovers accurate values of the GDD: $5480\,\textrm{as}^2$ from the spectrogram shown in Figure \ref{noisy_spectrograms}-a, and $5830\,\textrm{as}^2$ from the one in Figure \ref{noisy_spectrograms}-b. In comparison, FROG recovers GDD's of $6250\,\textrm{as}^2$ and $6100\,\textrm{as}^2$, respectively. This example demonstrates that ACE can tolerate very noisy spectra, and moreover that it is robust against errors in the vector potential $A_\mathrm{L}(t_0)$, as determined from the streaked spectra.

\section{Conclusion}

In conclusion, we have derived a general analytical expression (\ref{streaked_bandwidth}) for the change in spectral breadth due to the streaking effect by considering the trajectories of a photoelectron ejected by an isolated attosecond pulse in a laser field. We have used this equation as a basis for a method to directly extract the attosecond chirp from a sequence of streaked spectra. In contrast to the attosecond FROG retrieval, the ACE procedure does not require streaked spectra to be recorded with a delay step on the order of the attosecond pulse duration: it only requires the delay step to properly sample the streaking field. This alleviates many of the experimental constraints related to the current approaches to characterize isolated attosecond pulses. In addition, the ACE procedure is simple to implement, robust against experimental artifacts, and fast---taking seconds to execute and requiring very few ($\lesssim 10$) streaked spectra. This makes ACE ideal for real-time diagnostics in attosecond streaking measurements.

\begin{acknowledgement}
The authors are grateful for discussions with F. Krausz. This work was supported by the Max Planck Society and the DFG Cluster of Excellence: Munich Centre for Advanced Photonics (MAP). The final publication is available at www.springerlink.com
\end{acknowledgement}

\begin{appendix}
\section{A relation between duration, bandwidth and dispersion for arbitrary spectra}

The following is a proof of the general relation
\begin{align}
  \label{bandiwdth-dispersion-duration_relation}
  \tau^2&= \tau_0^2+\gamma^2\delta^2
\end{align}
between the duration $\tau$, the Fourier-limited duration $\tau_0$, the bandwidth $\delta$ and the group-delay dispersion (GDD) $\gamma$ for a pulse with an arbitrary spectrum and a constant GDD; $\tau$, $\tau_0$ and $\delta$ are taken as standard deviations of their respective distributions.

Let us first define spectral and temporal profiles as
\begin{subequations}
\begin{align}
  \label{spec_profile}
  \tilde{f}(\omega) &= \tilde{f}_0(\omega)\mathrm{e}^{\frac{\mathrm{i}}{2}\gamma\omega^2}\\
  \label{temp_profile}
  f(t) &= \frac{1}{\sqrt{2\pi}}\int_{-\infty}^{\infty}\tilde{f}(\omega)\mathrm{e}^{\mathrm{i}\omega t}\mathrm{d}\omega=\mathcal{F}^{-1}[\tilde{f}(\omega)](t).
  \end{align}
\end{subequations}
We assume, without lack of generality, that $\tilde{f}(\omega)$ and $f(t)$ are centered around $\omega=0$ and $t=0$, respectively.

The duration $\tau$ is defined as the standard deviation of $f(t)$, which is the square-root of the variance
\begin{align}
  \label{duration_tau}
  \tau^2&=\int_{-\infty}^{\infty}t^2|f(t)|^2\mathrm{d}t=\int_{-\infty}^{\infty}\left|-\mathrm{i}t f(t)\right|^2\mathrm{d}t\\
  \nonumber
  &=\int_{-\infty}^{\infty}\left|\mathcal{F}^{-1}[\tilde{f}'(\omega)](t)\right|^2\mathrm{d}t.
\end{align}
In the following derivation, the prime symbol (``$'$'') denotes differentiation with respect to the argument and the pulse is normalized according to $\int_{-\infty}^{\infty}|f(t)|^2\mathrm{d}t=\int_{-\infty}^{\infty}|f(\omega)|^2\mathrm{d}\omega=1$. Assuming $\gamma$ is frequency-independent, then (\ref{spec_profile}) implies $\tilde{f}'(\omega)=\tilde{f}_0'(\omega)\mathrm{e}^{\frac{\mathrm{i}}{2}\gamma\omega^2}+\mathrm{i}\gamma\tilde{f}(\omega)$. Inserting this expression for $\tilde{f}'(\omega)$ into the rightmost-hand-side of (\ref{duration_tau}), we obtain
\begin{subequations}
\begin{align}
  \label{duration_tau_2}
  \tau^2&=\int_{-\infty}^{\infty}|I(t;\gamma)|^2\mathrm{d}t+\gamma^2\int_{-\infty}^{\infty}|f'(t)|^2\mathrm{d}t\\
  \nonumber
  &+\gamma\int_{-\infty}^{\infty}\big(I^*(t;\gamma)f'(t)+I(t;\gamma){f'}^*(t)\big)\mathrm{d}t,\\
  \label{definition_of_I}
  I(t;\gamma)&=\mathcal{F}^{-1}[\tilde{f}'_0(\omega)\mathrm{e}^{\frac{\mathrm{i}}{2}\gamma\omega^2}](t).
\end{align}
\end{subequations}

In analogy to (\ref{duration_tau}), the bandwidth-limited duration is given by
\begin{align}
  \label{bandwidth-limited_duration_tau0}
  \tau_0^2&=\int_{-\infty}^{\infty}\left|\mathcal{F}^{-1}[\tilde{f}_0'(\omega)](t)\right|^2\mathrm{d}t.
\end{align}
Now, $I(t;\gamma)$ and $\tilde{f}'_0(\omega)\mathrm{e}^{\frac{\mathrm{i}}{2}\gamma\omega^2}$ are Fourier transforms of each other. Thus, from Parseval's theorem, we have
\begin{align}
  \label{Parseval_I}
  \int_{-\infty}^{\infty}|I(t;\gamma)|^2\mathrm{d}t&=\int_{-\infty}^{\infty}\left|\tilde{f}'_0(\omega)\mathrm{e}^{\frac{\mathrm{i}}{2}\gamma\omega^2}\right|^2\mathrm{d}\omega\\
  \nonumber
  &=\int_{-\infty}^{\infty}\left|\tilde{f}'_0(\omega)\right|^2\mathrm{d}\omega=\tau_0^2,
\end{align}
where (\ref{bandwidth-limited_duration_tau0}) in combination with Parseval's theorem was used for the last equation on the RHS of (\ref{Parseval_I}).

The bandwidth $\delta$ is given, also in analogy to (\ref{duration_tau}), as
\begin{align}
  \label{bandwidth_delta}
  \delta^2&=\int_{-\infty}^{\infty}\omega^2|\tilde{f}(\omega)|^2\mathrm{d}\omega=\int_{-\infty}^{\infty}\left|\mathrm{i}\omega\tilde{f}(\omega)\right|^2\mathrm{d}\omega.
\end{align}
Since $\mathrm{i}\omega\tilde{f}(\omega)$ and $f'(t)$ are Fourier transforms of each other, then from Parseval's theorem,
\begin{align}
  \label{bandwidth_delta_2}
  \delta^2&=\int_{-\infty}^{\infty}\left|\mathrm{i}\omega\tilde{f}(\omega)\right|^2\mathrm{d}\omega=\int_{-\infty}^{\infty}|f'(t)|^2\mathrm{d}t.
\end{align}

Using (\ref{Parseval_I}) and (\ref{bandwidth_delta_2}), we may now represent the duration $\tau$ as
\begin{subequations}
\begin{align}
  \label{duration_tau_3}
  \tau^2&=\tau_0^2+\gamma^2\delta^2+2\gamma\int_{-\infty}^{\infty}\mathfrak{R}[I^*(t;\gamma)f'(t)]\mathrm{d}t\\
  \label{duration_tau_4}
  &=\tau_0^2+\gamma^2\delta^2+2\gamma\int_{-\infty}^{\infty}\mathfrak{I}[\omega \tilde{f}'_0(\omega)\tilde{f}^*_0(\omega)]\mathrm{d}\omega,
\end{align}
\end{subequations}
where we have used $I(t;\gamma)=\mathcal{F}^{-1}[\tilde{f}'_0(\omega)\mathrm{e}^{\frac{\mathrm{i}}{2}\gamma\omega^2}](t)$ and $f'(t)=\mathcal{F}^{-1}[\mathrm{i}\omega\tilde{f}(\omega)](t)$ to obtain (\ref{duration_tau_4}). Now, if the pulse's GDD is constant over its spectrum, $\tilde{f}_0(\omega)$ is a strictly \emph{real} quantity, and therefore the last term on the RHS of (\ref{duration_tau_4}) is equal to zero, yielding (\ref{bandiwdth-dispersion-duration_relation}). $\square$

Relation (\ref{bandiwdth-dispersion-duration_relation}) is the reason why the ACE procedure can be applied for arbitrary XUV spectra (of course, provided that the attosecond pulse is short compared to the half-period of the streaking field).
\end{appendix}

%

\end{document}